# Orienting gaze toward a visual target:

# Neurophysiological synthesis with epistemological considerations

Laurent Goffart [1]*

1 : Aix Marseille Université, CNRS, Centre Gilles Gaston Granger, Aix-en-Provence, France

* : Correspondence: laurent.goffart@cnrs.fr

**Abstract**: The appearance of an object triggers an orienting gaze movement toward its location. The movement consists of a rapid rotation of the eyes, the saccade, which is accompanied by a head rotation if the target eccentricity exceeds the oculomotor range, by a slow eye movement if it moves. Completing a previous report, we explain the numerous points that lead to questioning the validity of a one-to-one correspondence relation between measured physical values of gaze or head orientation and neuronal activity. Conflating kinematic (or dynamic) numerical values with neurophysiological recordings carries the risk of believing that central neuron activity directly encodes gaze or head orientation rather than mediating changes in extraocular and neck muscle contraction. Rather than reducing mismatches between extrinsic physical parameters (such as position or velocity errors), eye and head movements are behavioral expressions of intrinsic processes that restore a poly-equilibrium, i.e., balances of activities opposing antagonistic visuomotor channels. Past results obtained in the cat and monkey left a treasure allowing a synthesis, which illustrates the tremendous complexity underlying the changes in the orientations of the eyes and head. Its aim is to serve as a guide for further investigations in marmosets or for comparison with other species.

# 1. Introduction

Among the numerous relations that an entire animal establishes with its environment, the orienting movement was once proposed to underlie this interrogative process which "*in Man has strongly developed in this highest form which is curiosity - the parent of that scientific method by which we hope one day to achieve a correct orientation in the knowledge of the world around us*" [1]. This view remains popular in the domain of contemporary cognitive neuroscience wherein orienting is a module underlying exploratory behavior, gathering sensory signals, transforming them into information and storing them in some sort of internal models of the external world.

In this article, we shall take a different point of view, avoiding teleological arguments as much as possible. We shall consider the orienting movement as the behavioral outcome of intrinsic brain processes, which are blind to the relations occurring at the macroscopic level, that is, between the entire animal organism and objects in the external world. From this standpoint, the inner milieu consists of multiple equilibria maintaining some kind of stability, and whenever asymmetrical changes disrupt it, they trigger intrinsic processes restoring the homeostasis. The restoration involves a coordinated set of reactions measurable at the musculoskeletal, vascular, vegetative and endocrine levels. Orienting movements of gaze, head or body toward a target are macroscopic, externally visible consequences of activities traveling through the sensorimotor neural networks during such restorations. Synaptic changes and the persistent traces left by repetition are internal consequences that are indirectly visible. In the same way as the sandy soil conforms to the cur-rent of a river while determining at the same time its flow, the sensorimotor neuronal channels assimilate to some extent the regularity of activity flows while constituting at the same time the set of constraints for the on-going flows.

The orienting movement is not the only type of reaction to an external stimulus. Other types include distancing, avoidance and escape. Orienting differs from them by the radically opposite way in which it manifests spatially in relation to the stimulus. Contrary to them, it consists of directing the body segment that carries the teleceptive organs towards the location of the stimulus, be it detected or coveted. In this article, we shall focus on the animal reaction toward a visual stimulus and on two of its physical characteristics: its velocity and its accuracy. Indeed, the orienting response consists neither of a sliding movement nor of a sequence of small movements directed toward the location of a targeted object. For the vast majority of animals, it manifests itself rather by a sudden and extremely rapid rotation of the mobile segment (body, head or eye) that carries the photosensitive epithelium. The amplitude and direction of the rotation are confined within a range of those sufficient to symmetrize back the functional neuronal image of the stimulus within the sensorimotor networks.

At the end of the orienting movement, measuring the amplitude of the angular dis-placement performed by the mobile segment gives a set of numerical values that are compared to the eccentricity of the target. Figure 1 illustrates their relationship in the goldfish [2]. The amplitude of the body rotation amplitude is not constant; it increases with the angular distance of the stimulus relative to the body. This relation has been documented in several other species: in the frog [3], the salamander [4], the barn owl [5], the cat [6], the macaque monkey [7-8] and the human subject [9-11]. Also observed in invertebrate animals such as insects (mantis: 12-13; dragonfly: 14) and crustaceans [15], it suggests that, beyond the diversity of photosensitive surfaces (retina or compound eyes), of muscle fibers, of their attachments - indirect or direct - to the mobile segment, the extension of physical space has somehow been embedded within the neural networks.

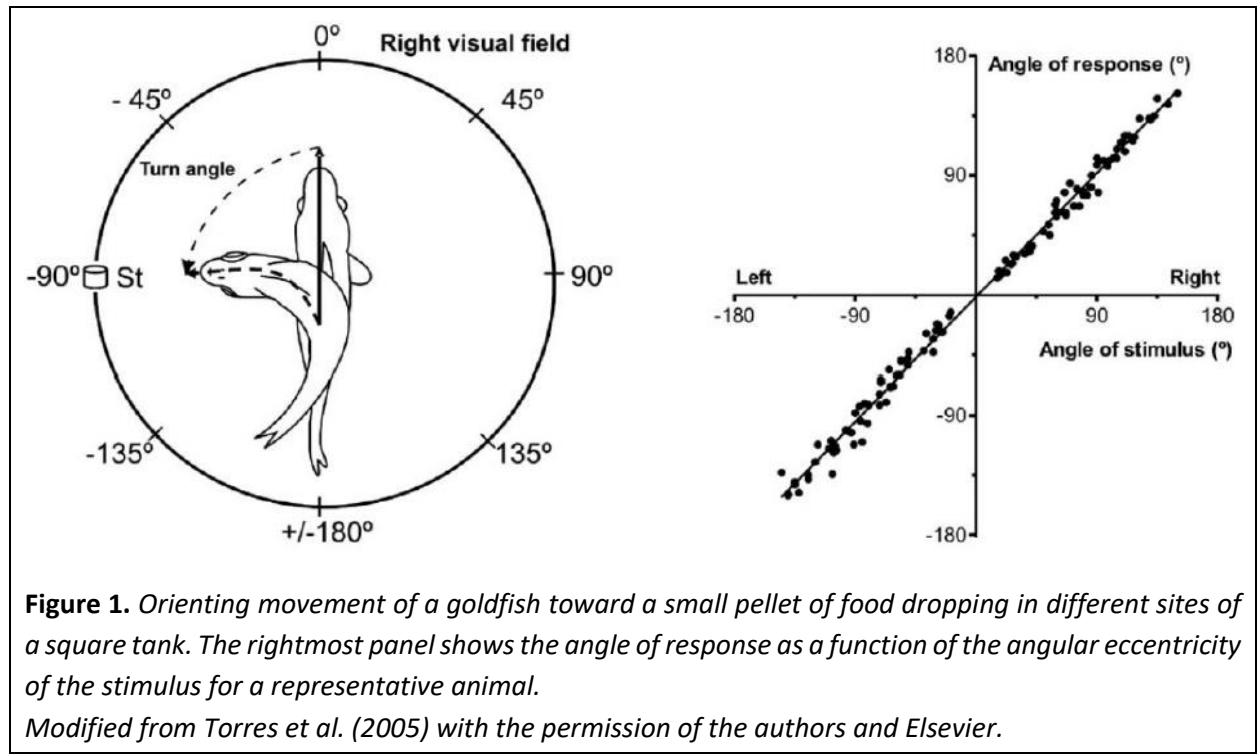

**Figure 1.** *Orienting movement of a goldfish toward a small pellet of food dropping in different sites of a square tank. The rightmost panel shows the angle of response as a function of the angular eccentricity of the stimulus for a representative animal.*
*Modified from Torres et al. (2005) with the permission of the authors and Elsevier.*

In the cat and monkey, the relation is not a fixed mapping between the topology of the photosensitive surface and a set of specific temporal commands changing the orientation of the eyes and head. Experiments in these animals showed that this behavioral invariant is robust to changes in saccade trajectory induced by a brief electrical stimulation in some territories in the brainstem (cat: [16-17]; monkey: [18-23]) or in the cerebral cortex (monkey: [23-24]). However, when the perturbation is applied in some sites of the pontine reticular formation [19] or in the medial cerebellum [25-26], the monkeys orient their gaze toward a location that is shifted with respect to the target location. They are shifted by an offset equal to the stimulation-induced change in eye position. This targeting error confirms that the extraocular muscle proprioception does not participate in guiding saccades toward their target [27].

## 2. Fundamental and epistemological issues

In response to the same initial physical conditions, the amplitude of the response is not constant but variable within some limits. Beyond the multiplicity of possible channels by which retinal signals can reach the motor neurons and beyond the time course of the persistence of signals, this variability, which is visible in Figure 1, is also due to the facts that objects in the external world are not points and that visual fixation does not involve the projection of their image onto one single light sensing cell, but several ones. In some species, the photosensitive cells are grouped within the fovea, a pitted invagination in the retina. We shall use the notion of "gaze" instead of the so-called "line of sight" to denote the imaginary "tube" through which photons travel, from the fixated object in the external world to the foveal photoreceptors.

Despite quasi-inevitable numerical differences, some degree of congruence must nevertheless be considered between the spatial location of the visual stimulus and the zone toward which gaze is directed. As mentioned above, this overlap is preserved even when the functioning of some key sensorimotor regions is perturbed. Before reporting these findings and discussing their implications, we shall address a few epistemological issues related to the risk of importing in the neurophysiological study of visual orienting, notions that are foreign to the brain functioning.

## 2.1. Distinguishing the medium of cerebral activity from physical space

The "binocular fixation point" is the intersection between two imaginary lines that characterize the direction of each visual axis while the "vergence angle" is the angle made by these two lines [28]. When that angle decreases, the eyes are said to converge and one speaks of convergence. When it increases, they are said to diverge (divergence). In the subject with no strabismus, when a saccade shifts gaze between two targets located along an isovergence curve, the spatial trajectory of the binocular fixation point follows neither the shortest physical path between the targets (the straight line) nor a path along the isovergence curve. Relative to the frontal plane, the binocular fixation point takes a path away from the isovergence curve followed by a return to it: a transient divergence happens during the saccade. The path traveled by gaze lies in a "geodesics" that does not belong to the zero-curvature space of Euclidean geometry. There is no force field in the external world that constrains gaze to take such a path. The path is intrinsically determined, caused by the neuronal networks upstream of the extraocular muscles.

As pointed out by Pellionisz and Llinás, "*since within the brain there is no 'instantaneous' simultaneity agent comparable to the light, the classical usage of separate space and time coordinates is inapplicable in the case of describing the inner workings of the central nervous system*" [29]. From a naive point of view, the timeless entity that geometers and physicists designate as "space", the receptacle occupied by objects, might correspond to the medium within which flows of neuronal activity propagate, from the excitation of sensory cells by external objects to the time when the orienting reaction ends. However, the cerebral networks radically differ from the three-dimensional framework that assigns numerical values to positions and displacements of objects. They constitute a medium that is neither isotropic nor homogeneous, but composed of multiple and diverse channels. Each physical place captured by gaze is the outcome of multiple flows of activity through this medium. The divergence characterizing the afferent pathways first broadens the neural image of a small spot of light, an image which is then compressed by the convergence of activities toward the motor nerves. Transforming the recruitment of excitatory and inhibitory synapses and the firing rate of neurons to algebraic time functions engenders a continuum that is not in accordance with the remote and distributed conditions of spike triggering.

The multiple sensorimotor channels cannot correspond to a neural representation (or correlate) of physical "space" for other reasons. The numerical values assigned to different physical target positions or movement endpoints are independent of each other. They are ordered and aggregate to each other within another continuum, which corresponds to an infinite number of simultaneous points. In the brain, there is no corresponding entity. The elements are neither compact nor rigid, and their number is much smaller than the number of objects present in the external environment of a subject. Activities in the brain corresponding to different physical locations are neither points nor lines. Those leading to the generation of movements towards nearby but different locations can involve common neurons. Indeed, neurons have extended and overlapping response fields [30]. The sharing of neuronal elements is such that two simultaneous visual targets can lead to a saccade (called averaging saccade), which directs gaze toward a location situated midway between the two targets or along the line connecting them [31-32].

Moreover, whatever corresponds to physical space in the brain, it is not permanent, but short-lived. Its persistence requires the presence of objects. When we consider the ac-curacy and precision with which saccades indicate the location of a target, they depend considerably on its presence. When it is produced in the very near future after target appearance, the saccade is sufficiently accurate to confirm the presence of the object by a set of sensations, which are relatively always the same and characterize the foveation. Regarding the object location, the kind of "measurement" made by saccades is, so to speak, reliable. By contrast, when the saccade is made after a delay, even a relatively short one, the "measurement" is no longer reliable [33-35]. Visual sensations can no longer confirm the presence of the object since it has disappeared while the change in saccade accuracy and precision make the movement-related sensations uncertain. In other words, it is as if the persistence of visual signals in the brain required the presence of objects to create a "space", in which case its internal representation (or neuronal reproduction) may not necessary since it is already out there.

Finally, the neural plasticity, the irreversibility of activities and their canalization argue for a fundamental difference between the cerebral medium and the Newtonian receptacle. Yet, the adaptation of movements to environmental constraints led some authors to propose that the brain activity emulates internal models of the physical laws of motion. Unfortunately, they missed to explain how the diversity of animals and nervous systems is compatible with the paucity of the laws of motion. Leaving aside the principle according to which rest corresponds to a balance of forces that cancel each other out, they did not explain either how the brain activity could sustain at constant speed the motion of a body segment with muscles of finite length. If the laws of mechanics are conventions [36-37] used to quantify the movement of objects in space, then it draws into question whether it is reasonable to assume their applicability to the brain sciences. The consistency and the viscosity of the cerebral medium are not comparable to those of the environments (water, ground and air) in which the animal body can move. As for the eye movements, their study in the untrained monkey shows a tracking that is not as smooth as the continuous and uniform motion of the visual target. The visual tracking is discontinuous and saltatory, interrupted by saccades [38-40] indicating that the streams of visuomotor activity do not spontaneously mimic the constant motion of the target. The fact that repeated practice is required to perform spatially and temporally accurate movements does not imply cerebralization of physical laws. It merely illustrates the flexibility of the brain's functioning to perform accurate movements.

Refusing to import notions pertaining to the kinematic or kinetic description of movements onto the activity of neurons does not lead the neurophysiology of eye-head movements to a dead end. We shall see that it provides the opportunity to take a fresh start and to investigate the brain functioning differently from the way that was promoted during the recent decades.

## 2.2. The spatiotemporal transformation

A fundamental challenge in the brain sciences is to identify and extract processes from the complex networks of polyvalent cells situated between the retinal ganglion cells and the extraocular and neck muscle fibers. Because of the involvement of neurons in multiple tasks, the delineation of processes is far from being a trivial task. The mere excitation of a small group of ganglion cells in the retina by a stimulus as simple as a brief spot of light is transmitted to a multitude of post-synaptic neurons in several deep cerebral nuclei (e.g., lateral geniculate nucleus, superior colliculus, nucleus of the optic tract). The spikes emitted by these neurons are then transmitted to more numerous post-synaptic cells distributed in multiple areas of the cerebral cortex. The spot of light elicits a neural activation that spreads out within the brain networks. Its expansion is associated with a temporal persistence insofar as the duration of elicited activities outlasts that of the stim-ulus. This maintained activity is necessary for changing the firing rate of specific groups of motor neurons and for driving the contraction of extraocular and neck muscles until gaze is oriented toward the location of the spot of light, sometimes a long time after its disappearance.

The process by which different loci and more distant stimuli in the visual field (i.e., more eccentric loci of retinal activity relative to the fovea) lead to different and longer motoneuronal discharges has been called "spatiotemporal transformation". Such a trans-formation is not limited to orienting movements toward a visual target. It characterizes all movements directed toward events that are spatially delimited such as an object in the visual field or a point stimulation on the surface of the skin, as well as toward events whose physical location is more diffuse such as sounds or screams. In all cases, a premature arrest of motoneuronal activities will make the movement hypometric, i.e., falling short of its target.

Initially, the dysmetria of movements made by patients suffering from cerebellar dysfunction led to propose that the cerebellum performed this spatiotemporal transformation [41-42]. The afferent signals from retinotopically organized maps such as the one found in the superior colliculus were thought to be converted into a specific temporal pat-tern of motor activity (see Fig. 10 in [43]). However, the observation of accurate saccades in patients suffering from spinocerebellar degeneration, despite dramatic slowing and increased duration, led to abandon this hypothesis [44]. The conjecture was then proposed that saccade duration be determined by the work of a negative feedback loop [44-46]. The premotor neurons would be driven by a motor error command that results from the continuous comparison between two signals: a reference signal specifying the goal or desired state (desired eye orientation or desired change in eye orientation), and a feedback signal estimating the current state (current eye orientation or actual change in eye orientation). As long as a mismatch persists between these two signals, the motor error command drives the premotor neurons until the estimated current state matches the desired state. In this theoretical framework, there is no signal specifying the duration of saccades; it merely corresponds to the time taken by a feedback signal to zero-out the motor error command. Saccade inaccuracy would result either from an altered reference signal or from impaired feedback signal. Transient changes in the trajectory of saccades should not affect their ac-curacy as long as it does not alter the signal that specifies the saccade goal, or the amplitude and direction required to foveate the target [17, 19, 21].

For several years, the negative feedback loop remained the simplest and widely shared hypothesis to account for the spatiotemporal transformation. The reference signal was shown to be conveyed by the population of active neurons in the intermediate and deep layers of the superior colliculus [47-49] while the comparison with feedback signals took place downstream in the reticular formation, under the influence of neurons in the midline cerebellum [25].

Figure 2 illustrates how the direction (A) and amplitude (B) of saccades toward a visual target are preserved in spite of their dramatic slowing when the activity of neurons in the deep superior colliculus is suppressed by the local injection of a blocker of Na+ membrane channels [50]. After each injection of a small amount of lidocaine, the peak velocity of saccades is greatly reduced (C) while their duration is increased (D) by an amount that appears proportional to the amount of velocity reduction. In this experiment, the injection did not silence all collicular neurons, but only cells that fired maximally for saccades having particular direction and amplitude. Saccades with amplitude and direction that were close to this optimal vector were inaccurate [48]. In addition to demonstrating that saccade metrics was specified by the population of active neurons in the superior colliculus [47], this study brought evidence for a causal link between collicular activity and saccade velocity and for a feedback control of saccade amplitude and direction (see also 51).

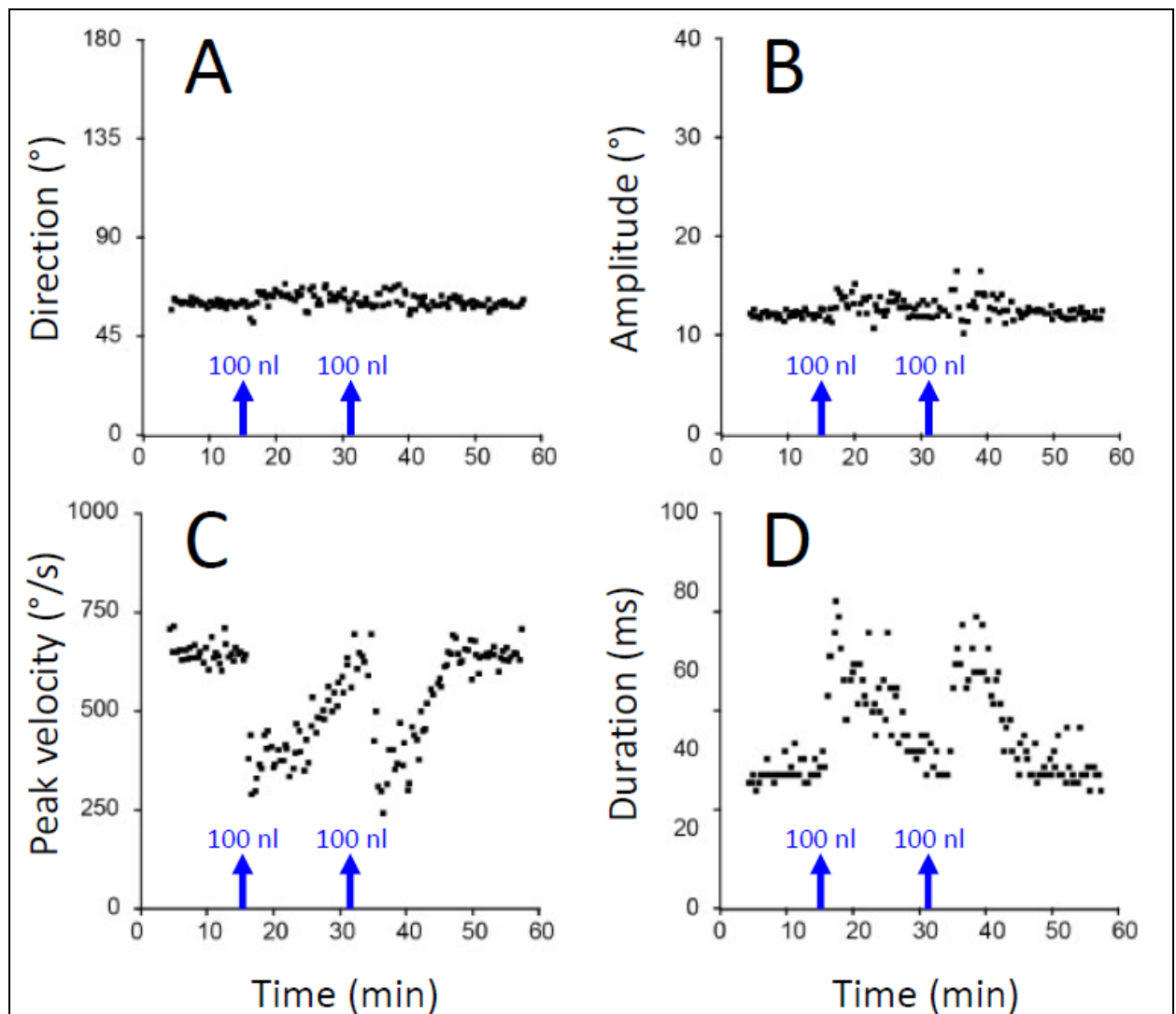

**Figure 2.** *Effects of suppressing the activity of a small set of collicular neurons on saccades toward a visual target. A small volume of lidocaine (100 nl) was injected two times inside the population of active neurons while a monkey made saccades toward a target. Courtesy of Dr David L. Sparks, modified with his permission.*

Around the same time, works in the head unrestrained cat led to propose that the comparison between the reference and feedback signals was performed within the superior colliculus itself, by a caudo-rostral change in the locus of activity [52-53]. This topo-graphical change of collicular activity would switch from the recruitment of saccade-related neurons to the recruitment of so-called "fixation" cells. The former neurons would be distributed across the entire extent of the SC whereas the latter would be located in the rostral part. The saccade would stop as the "fixation" cells resume their sustained pre-saccadic firing rate, which in turn, would excite cells located in the nucleus raphe interpositus (RIP). Therein, neurons called omnipause neurons (OPN) exhibit a sustained spontaneous firing rate and pause during each saccade, regardless of its amplitude and direction. Their sustained activity was proposed to inhibit the premotor burst neurons and their pause to trigger the saccade. The conjecture of caudo-rostral change in the locus of activity was abandoned after the publication of studies reporting the failure to observe any migration of activity across the SC during saccades [54-57, but see 58].

Later, in line with observations reported in an earlier study [59], SC neurons were shown to emit a burst of action potentials during miniature saccades [60-61], extending the saccade-related function to the rostral SC, as a microstimulation study suggested [21]. Indeed, contrary to electrical microstimulation of OPN, which only decelerates and interrupts visual saccades, microstimulation of the rostral SC deviates their trajectory. The absence of hypermetria after pharmacological inactivation of the rostral SC [62-63] also questioned its role in terminating saccades whereas the reduced rate of microsaccades confirmed its involvement in the generation of miniature saccades [59-61, 63-64]. Finally, the lack of impairment in either the latency or the accuracy of saccades after RIP lesion questioned even the involvement of OPN in fixation and saccade termination [65-66].

Altogether, the set of these numerous studies and others led to realize that the spatiotemporal transformation might not be explicitly performed by distinct groups of neurons, which neurophysiologists ought to search and identify within the brain. Instead, the transformation would be supported by activities of populations of neurons that are massively interconnected and distributed over several neural territories [67]. We shall now see that the same conclusion holds for the cerebral control of gaze direction during visual fixation.

## 2.3. Gaze direction as a poly-equilibrium

Several lines of evidence led to propose an alternative to the "fixation-saccade" dichotomy initially proposed by Munoz and Guitton [68] and still defended nowadays by an electro-anatomical study [69]. Firstly, as mentioned above, the absence of fixation instability during lesion or during pharmacological inactivation of RIP [65-66] questions the necessary involvement of omnipause neurons in maintaining gaze stable. Secondly, several studies made in awake animals report cases in which the animal does not direct its gaze toward a visual target during fixation, or does not correctly direct its head toward a food target. During unilateral inactivation of the caudal fastigial nucleus (cFN), cats direct their gaze and mouth toward a location that is offset with respect to the physical target location [6]. A comparable eccentric fixation happens in the head restrained monkey during unilateral pharmacological inactivation of cFN. Therein, the local injection of an inhibitory agent (muscimol) leads to an ipsilesional fixation offset [6, 70-73] whereas the injection of a disinhibitory agent (bicuculline) yields a contralesional offset [74]. The muscimol-induced offset is not a defect in positioning the eyes in the orbit, but a gaze-related disorder because its size is similar between the cases in which visual fixation is made with the head restrained and unrestrained. When the head is free to move, a large ipsilesional deviation affects its orientation and the deviation of the eyes in the orbits is contralesional [75]. A weakening of fixation stability accompanies this offset insofar as mon-keys have difficulty to maintain their gaze near the central target when other behaviorally relevant targets appear in the periphery. Irrepressible saccades are made toward them during delayed oculomotor tasks (our unpublished observations; see also the increased rate of no-go errors in [76]).

Anatomical studies report that cFN neurons project neither to the nucleus prepositus hypoglossi nor to the medial vestibular nucleus [77] where tonic eye position neurons are located [78]. However, they project to the rostral part of both superior colliculi [79]. Thus, asymmetrical sustained firing rate in cFN would affect the balance of bilateral activity in rostral SC whenever gaze is directed toward a target in the central visual field. This conjecture is confirmed by the consequence of unilaterally inactivating small sectors in the rostral SC. After the local injection of small amounts of muscimol, monkeys exhibit an offset when they fixate a target, regardless of whether it is static [63] or moving [80]. They also make irrepressible saccades toward the peripheral targets during the memory-guided saccade task [62]. Asymmetrical sustained firing rate in cFN and rostral SC is also trans-mitted to pre-cerebellar pontine nuclei when the target is moving [81]. As for the ipsilesional deviation of the head [6, 75], it would result from asymmetrical input to reticulospinal neurons [82].

Difficulties in directing the eyes toward the location of visual target have also been reported during unilateral inactivation of small part of the frontal eye fields [83]. The observation of a fixation offset is not restricted to lesional cases. An upward fixation offset is observed in the scatter of gaze direction when a macaque monkey looks at a small visual target in the dark [84-86]. Asymmetrical representations of the upper and lower visual fields in macaques [87-88] can account for this offset as well as for the upward bias of horizontal saccades [89]. While the population of active neurons in the SC is immune of this asymmetry in darkness, when a visual target is presented in an illuminated background, the active population may be different. More cells sensitive to the lower visual field may be recruited to counteract the upward bias caused by the asymmetrical representation of the upper and lower visual fields in the SC [88]. In darkness, the absence of their recruitment would account for the upward bias of horizontal saccades.

Altogether, these results led to revise the view of fixation as a process inhibiting the generation of saccades [62, 68] and to propose the conjecture of fixation as an equilibrium [63, 89-90], later revised as a "poly-equilibrium" [91].

According to this alternative theory (Figure 3), a saccade or a slow eye movement is not initiated as long as the visuo-oculomotor system is within a mode where opposing commands (issued for instance by the left and right parts of the brainstem) counterbalance each other. For generating saccadic and pursuit eye movements, the symmetry breaking involves different groups of neurons. Saccades involve the recruitment of burst neurons located in the pontomedullary reticular formation whereas slow eye movements involve the recruitment of neurons in both nuclei of the optic tract, pontine nuclei and thalamus. Prior to generating a saccadic or slow eye movement, the target-related signals travel toward the motor neurons, across networks composed of cells that evince a spontaneous sustained firing rate, which maintains the stationarity of the eyes and head. This set of tonic premotor activities form an equilibrium of commands that counterbalance each other.

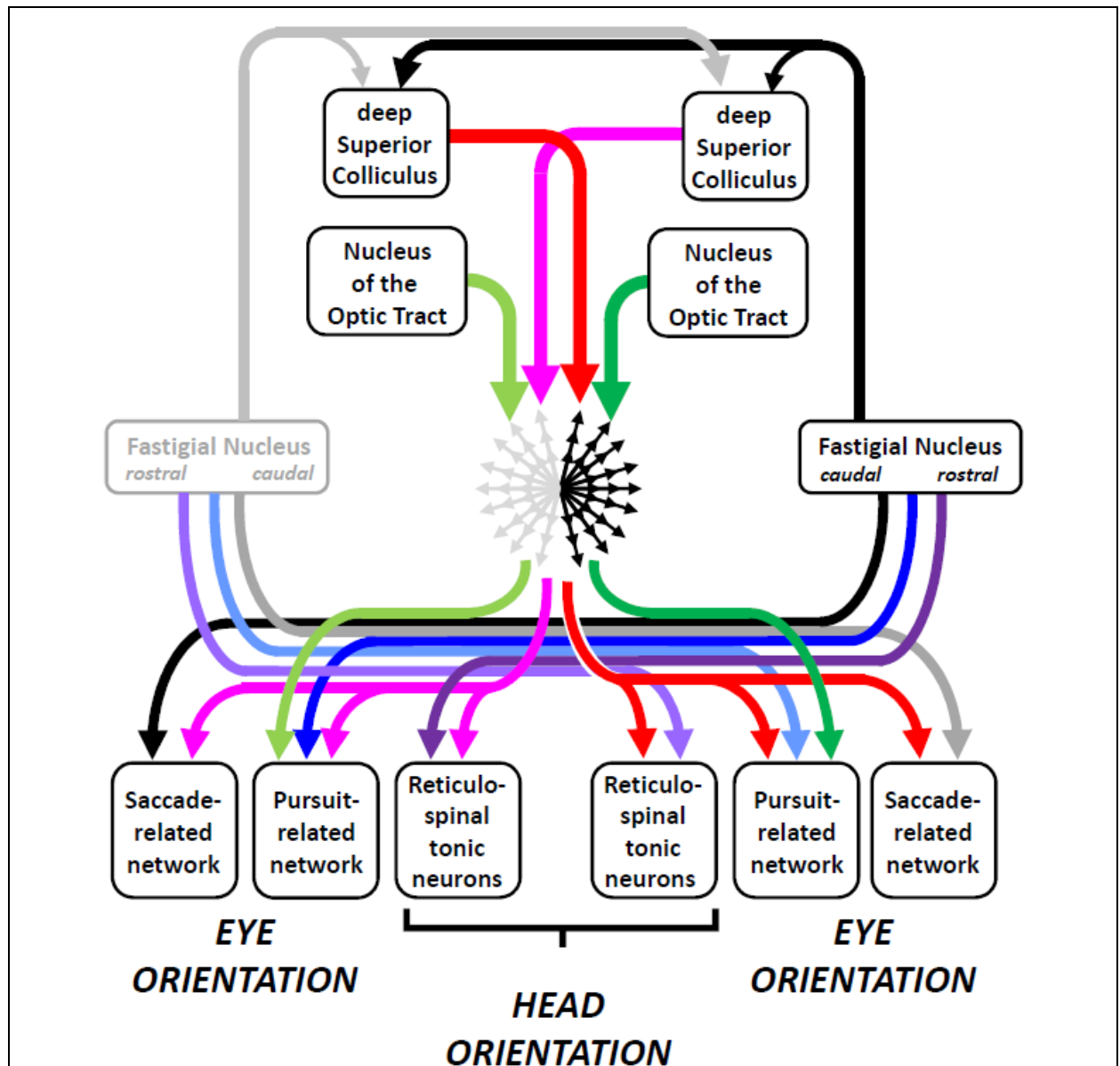

**Figure 3**. *Poly-equilibrium hypothesis: A saccade or a slow eye movement is not initiated when the visuo-oculomotor system is within a mode where opposing commands (issued for instance by the left and right superior colliculi or by the left and right nuclei of the optic tract) counterbalance each other. For generating saccadic and pursuit eye movements, the symmetry breaking involves different groups of neurons. Saccades involve neurons located in the pontomedullary reticular formation whereas slow eye movements involve neurons in the nuclei of the optic tract, the pontine nuclei and the thalamus. The bilateral fastigial activity also contributes to the neck muscle tone, which specifies the horizontal orientation (yaw) of the head through fastigio-reticular projections.*

The notion of poly-equilibrium is more appropriate for naming this balanced activity because the equilibrium is multiple. It involves spikes conveyed through the channels that lead to the generation of saccades, the channels that lead to the generation of slow eye movements, those yielding the near response (accommodation and vergence) and those holding the atlas of the head [91]. The change in head orientation during unilateral lesion of caudal fastigial nucleus [6, 75] or superior colliculus [92] indicates that bilateral activity also contributes to the muscle tone that specifies the static orientation (yaw) of the head. Sustained activities are distributed within different groups of neurons in the left and right parts of the reticular formation (including pontine long-lead burst neurons and pursuit-related neurons), the cerebellum, the deep superior colliculi and the cerebral cortex. Concerning the deep superior colliculi, the activity involves their rostral pole but it is not restricted to them. It extends more caudally, even during steady fixation. Indeed, during the gap interval, saccade-related neurons in both SC fire, even those located at sites that are sensitive to target eccentricities of 8 deg (see Fig. 4 in [93]). Any suppression of activities within the set of commands participating to this poly-equilibrium alters the direction of gaze while fixating a target if it is not counterbalanced. As mentioned earlier, gaze is directed toward a location that is offset relative to the target, regardless of whether it is static [63, 70-75, 83] or moving [80-81]. Finally, unbalanced activity between the left and right superior colliculi promotes the generation of saccades toward the side opposite to the most active SC. By contrast, unbalanced activity between the left and right nuclei of the optic tract (NOT) promotes a slow eye movement toward the side to the most active NOT [89-91, 94].

## 2.4. The wanderings of the brain machine

At the time when bioengineering approaches started to guide the neurophysiological inquiry of eye movements with systems analytical techniques, David A. Robinson already warned that "*block diagrams of oculomotor organization serve as a compact description of system behavior but seldom have much bearing on the way in which the real system, composed of nerve and muscle, actually operates. The models thus do not contribute much to the neurophysiology (or neurology) of eye movements and incur the danger of suggesting that there actually are segregated portions of the nervous system which perform the differentiation, integration and other operations indicated in the boxes of the diagrams*" [95]. Neuroanatomical and neurophysiological studies reveal indeed functional properties and connectivity that are not visible in cybernetic diagrams. Establishing correspondences between cybernetic diagrams and the brain is analog to comparing an automaton to a living animal. It led some authors to search for various sources of noise within the brain activity rather than for differences between brains and machines. Contrary to a machine, recruiting the same group of photosensitive cells does not lead to the same response.

In spite of evidence for reliable saccade motor commands [96], some authors posited their variability [97] and even various kinds of noise [98] to explain the cerebellar control of saccade precision. The initial testing conditions, such as the starting eye position and its determining importance in specifying the locus of activity on the retina, the different brain connectivity between subjects were discarded, as were the short-lived character of visual signals, and the inevitably variable distributions of spikes within the massive plexus of neurons interposed between the spot of retinal activity and the last muscle fiber twitch.

Hence, another concern of bioengineering approaches was the incitation to coordinate (in the metrological sense) the set of numerical values brought by behavioral measurements and neuronal activities. While the target location and the location toward which gaze is directed are point values, their difference a segment (called error) and the movement trajectory a curve, their corresponding correlates in the brain are flows of activity, i.e., dynamic entities. As explained earlier, the projection of a mere brief spot of light on the retina leads to a pattern of activity that is organized as a plexus. Another brief spot of light exciting a different retinal region yields to another plexus. Even in the absence of overlap on the retina, the two plexuses involve common neuronal elements if the spots of light exincite neighboring groups of ganglion cells. Response fields of central neurons that overlap are the main evidence for this sharing of neuronal elements. This situation does not lead to confusion if "the precision or accuracy of a saccade results from the summation of the movement tendencies produced by the population of neurons rather than the discharge of a small number of finely tuned neurons" [45]. The effects of variable discharge of neurons can be reduced by averaging the activity over a population of many neurons, reducing further the correlation between the firing rate of single neuron and eye movement kinematics (or dynamics).

# 3. Orienting gaze toward a moving target

Many objects in the world are moving and orienting movements are also made toward them. In primates, their appearance in the peripheral visual field elicits a saccade that brings the target image onto the foveae. Then, the foveation is maintained by slow pursuit eye movements interrupted by catch-up saccades. Movement accuracy consists of directing gaze toward the location where the target is here and now, the target location at the saccade landing time. The temporal interval between the onset of sensory signals and the landing time led some authors to defend a view according to which the functioning brain would be endowed with so-called "predictive" capabilities. In this section, we shall explain that this inference is the consequence of conflating physical events and numerical values attributed to them with their physiological correspondence.

## 3.1. The interceptive saccade

Benefiting from high resolution recording of eye movements, studies showed that human and non-human primates manage to direct their gaze toward the actual location of a small target moving at constant speed [99-102] with an accuracy that is quasi-similar to that of saccades toward a static target [103]. This ability to intercept with gaze a small moving target seems to depend neither upon prior experience [104] nor upon the occurrence frequency of path taken by the target [105]. It occurs even when prior to its onset the saccade is perturbed by an unexpected change in eye position [106]. However, saccades toward a moving target are not always accurate. The accuracy is lost when the target does not move at constant speed: saccades to an accelerating target fall short of it whereas those to a decelerating target overshoot it [107].

The ability to bring the image of a moving target within the central visual field led some authors to suggest an ability to foresee (or "predict") the location that the target will occupy at the time of saccade landing. Indeed, because of the polysynaptic path between the retinal ganglion cells and the extraocular muscles, the signals that drive the firing rate of motoneurons cannot originate from the retinal site that the target is crossing presently. They originate from a site that was crossed several tens of milliseconds earlier. The delay of visuomotor transmission is not known, but it likely depends upon the number of active neurons, upon their firing rate and the simultaneity of presynaptic action potentials at each relay [108]. As for saccades toward a static target [109-110], the latency of interceptive saccades may also depend upon the target contrast.

The study of saccades toward a transient target moving with a constant speed re-ports that several saccades landed at locations that the target did not cross [107]. Many landed beyond the location where the target disappeared, along the path that it would have taken if it had remained visible. Because the monkeys had no prior experience with such small moving targets, these saccades could not be guided by memory-related signals. They were guided by signals that persisted after the target disappearance. However, the persistence is rather limited because the kinetics of the retinal streak (the time course of the target motion in the visual field) is taken into account. For identical paths, the pattern of landing positions depends upon the rate of change in target speed. As mentioned earlier, the saccades are accurate if the target moves at constant speed; they overshoot its current location if it decelerates, fall short if the target accelerates. It is not known whether accelerating and decelerating targets recruit different neuronal plexuses, whether the different saccade landing positions result from different spatiotemporal patterns of activity.

Lesional studies show that the impairment of some brain regions can affect saccades toward a moving target without altering those aimed at a static target [111-113]. This dissociation indicates that static and moving targets involve different sets of neurons. As for curved saccades made in response to a target that steps from one location to another (double step saccades [114]), the recruitment of the same neuronal plexus for accelerating and decelerating targets would be another evidence for the importance of considering the distribution of activity within the neuronal networks for the causal determination of behavioral responses. Although important, the knowledge of anatomical connectivity is not sufficient.

### 3.2. The post-saccadic slow eye movement (pursuit)

A slow eye movement follows the interceptive saccade. Studies in the monkey revealed that its speed does not spontaneously match with the target speed and that it increases with the animal's training and prior experience [104, 115]. Comparable observations were made in children [116-117]. In response to a continuous motion, naive mon-keys and children track a target with a succession of catchup saccades interspersed by intervals during which gaze direction lags behind the target with a distance that increases with time until the next catchup saccade. The slow movement that follows the interceptive saccade is classically called pursuit eye movement.

Clinical studies in human subjects [118-121] and in lesioned macaques [111, 122-125] indicate that saccadic and pursuit eye movements recruit different sets of neurons. Indeed, some lesions impair slow pursuit eye movements without altering saccades, and vice-versa. The segregation of saccade-related and pursuit-related networks is confirmed by: i) developmental studies showing different maturation speeds of saccadic and pursuit performances [116, 126-128], ii) separate adaptive adjustments of saccadic and pursuit eye movements [129-131], iii) imaging studies in healthy human subjects [132-133], and iv) the electrophysiological identification of saccade-related and pursuit-related neurons in distinct regions of the pontine tegmentum [134-136] and cerebral cortex [137-143]. In the cerebral cortex, the saccade-related subregions are preferentially interconnected with each other, and likewise for the pursuit-related subregions [143]. In summary, from the cerebral cortex to the reticular formation, the networks involved in the generation of pursuit and saccadic eye movements are relatively independent and parallel.

This parallelism contrasts with the claim that a "single sensorimotor process" drives the generation of both movement categories [144-146], a conjecture that is mostly based on behavioral observations and on a mistaken interpretation of neurophysiological observations. Firstly, the fact that neurons in rostral SC modulate their firing rate during both miniature saccades and pursuit [147-151] is the signature of looking straight ahead, i.e. moving gaze neither rightward nor leftward, neither upward nor downward, but "in all directions at the same time". As explained earlier, bilateral rostral SC activity maintains the equilibrium necessary to maintain gaze direction toward the target, regardless of whether it is static or moving, actual or virtual. We conjecture that this bilateral activity concerns also the nuclei of the optic tract and their post-synaptic targets involved in the generation of slow eye movements [89-91, 94]. Transmitted to pontine nuclei, bilateral activity is maintained through distinct loops between the cerebellum and the brainstem. Adjusting the functional symmetry that makes each equilibrium is required to compensate for unequal number of synapses and cells. Secondly, it is premature to conclude an involvement of omnipause neurons in pursuit eye movements because their electrical microstimulation slows them [152-153]. The slowing could result from inhibiting motor neurons by the retrograde [154-155] excitation of synapses from inhibitory burst neurons. Thirdly, the observation that sub-threshold electrical microstimulation in the lobules VIc-VII of the cerebellar vermis perturbs both saccadic and pursuit eye movement [156] remains compatible with a segregation of saccade-related and pursuit-related neurons. Therein, Purkinje cells firing during both saccades and pursuit are not numerous (approximately 10% [157]). Based on a large sample of neurons, another study reported that only 4% of saccade-related Purkinje cells modulated their firing rate during pursuit [158]. More recently, the direction tuning of Purkinje cells was found to differ between saccadic and pursuit eye movements, suggesting an independent processing of their spikes in the fastigial nuclei [159]. In the latter, approximately one third of the neurons emit a burst of action potentials during saccades and modulate their firing rate during pursuit eye movements [160]. However, emitting a burst of spikes does not necessarily imply that their post-synaptic target belong to the saccade premotor network. The burst of "saccade-pursuit" neurons may be transmitted to pre-cerebellar pursuit-related pontine nuclei and contribute to the post-saccadic enhancement of slow eye movements [161]. Finally, unilateral inactivation of cFN leads to severe disorders in saccadic and pursuit eye movements that are not correlated [81], whereas bilateral lesion leads to bilateral hypermetria of saccades [70] while the gain of pursuit eye movements is either reduced [162] or unchanged [163].

In conclusion, behavioral observations provide helpful information for guiding neurophysiological investigations when they take into account the anatomo-physiological constraints. Before giving several examples illustrating this benefit, we shall discuss about the capacity to foresee the invisible and to "predict the future".

### 3.3. About the "predictive" power of the brain

Because they direct gaze toward locations that have not yet been crossed by the tar-get, interceptive saccades have been called "predictive" and scenarios were proposed to explain foresight. For instance, the visuo-oculomotor system would be endowed of either a temporal countdown estimating of "the time remaining until the collision of the target with the line of sight" [164] or a predictive clock announcing "the time at which the eye trajectory will cross the target" [165]. This conjecture led to imagine unrealistic options such as "a spatial lead of the gaze at the saccade end, instead of attempting a precise capture of the target" [164, 166]. Indeed, the overshoot of saccades is incompatible with most published data. The majority of studies report that interceptive saccades either fall short, or land approximately at the location where the target is at their landing time [99-105, 160], even when their trajectory is perturbed by a change in eye position elicited by a brief electrical microstimulation in the SC [106]. Recent studies in the monkey showed that knowing the target path in advance does not lead to an overshoot of saccades either. Neither the accuracy nor the precision of saccades differ between blocks of trials in which the target path is always the same for one trial to the next and blocks in which the target path varies across the trials [105]. Saccades landing beyond the target location may occur in human subjects, but their occurrence is rare and conditioned by specific task instructions. Slightly more frequent when the target moves at constant velocity along a periodic path, they are almost eliminated when the subject is instructed to look at the target for detecting a change in its shape [168].

At the neuronal level, when a saccade is made toward a target moving at constant speed, saccade-related neurons in the deep SC emit a burst of action potentials as during saccades toward a static target. However, the population of active neurons does not contain commands related to future locations of a moving target. No spike is emitted by cells whose response field correspond to upcoming saccadic vectors [169]. The active population consists of a continuum of cells ranging from neurons issuing commands related to past locations of the target to neurons issuing commands related to the current target location. The topography of active neurons does not change as the target in the visual field for residual activities related to recently traveled locations persist.

The same conclusion holds for the results obtained in a preliminary study [170]. When comparing the response field during saccades toward a static versus a moving tar-get, a shift was found in the amplitude associated with the maximum firing rate. Cells exhibited a maximum discharge during saccades of larger amplitude when they were made toward a centrifugal target [see also 169]. However, in that study, the target moved so fast (60°/s) that it was possible that gaze did not capture it, unless the saccade was anticipatory. Indeed, a target that moves away from the central visual field with a speed of 60°/s after an initial jump of 2 to 6° reaches an eccentricity of 14 to 18° two hundred milliseconds later (approximate duration of response time). Since most measurements illustrated in their representative figure correspond to saccades smaller than 14 degrees, one is led to deduce either that the saccades were anticipatory, or that their amplitude was strongly hypometric. If they were hypometric, then the need of an additional compensatory command to ensure saccade accuracy becomes unclear. If the saccades were neither hypometric nor anticipatory, then the recorded bursts ought to correspond to that elicited by a past position of the target. Thus, the size of the shift in maximum firing rate between saccades to static and moving targets should be larger as the activity corresponds to positions of the moving target further in the past. Current knowledge does not allow saying whether a horizontal diffusion of neuronal activity happens at the level of the SC or at the level of collicular afferent input from structures of the cerebral cortex, such as LIP area [171-172] or FEF [173-174].

## 4. Reviving strictly neurophysiological explanations

Conflating kinematic (or dynamic) numerical values with neurophysiological recordings carries the risk of believing that central neuron activity directly encodes gaze or head orientation rather than mediating changes in extraocular and neck muscle contraction. Some studies examined how the firing rate of some central neurons could be ac-counted for by a combination of kinematic parameters such as position, velocity and acceleration of the eyes [175-178], sometimes without separating the cells whose firing rate precedes the combination from those whose firing rate follows it [179-180]. When one considers i) the spatial and temporal transformations that a brief spot of retinal excitation undergoes on its way to the motor neurons, ii) the temporal overlap of activities corresponding to successive target positions, iii) the multiple possible patterns of muscle contraction specifying any static orientation of gaze, and iv) the multiplicity of possible neuronal drives, one is led to doubt about a neurophysiological correspondence in the brain activity of notions such as dynamic position error, velocity error or even velocity.

Physical parameters are certainly convenient for interpreting and comparing the activity of neurons, sorting them in different groups and promoting scientific debates. How-ever, the reservations that David A. Robinson made about the block diagrams of oculomotor organization should not be restricted to the operations (differentiation, integration, gain …) performed by the blocks [95]. They also concern the signals proposed to be "encoded" in the firing rate of neurons such as instantaneous motor error, velocity error or even velocity. The motor error command that feeds the premotor neurons during saccades toward a static target [147, 181-182] might be less problematic because it is an oculocentric command [183] that can result from the preserved retinotopic arrangement of fibers connecting different neuronal assemblies.

To sum up, questioning the validity of a one-to-one correspondence relation between the sequence of values measured at the macroscopic level and the activity of microscopic neuronal elements does not lead the neurophysiology of gaze orientation to an epistemological dead end. An alternative approach remains possible. Rather than reducing mismatches between extrinsic physical parameters (such as position error or velocity error), eye and head movements can be viewed as the behavioral expression of intrinsic process-es that restore balances of activities opposing antagonistic visuomotor channels.

Interpreting the firing rate of neurons with notions belonging to the physical description of the movement of a body segment should be completed by the identification of muscle groups and neuronal networks involved. Comparative studies of the consequences of their functional perturbation, clinical and neuroanatomical studies should remain a privileged guide for interpreting neuronal activities and understanding their contribution to behavioral performances [184-185]. Building upon the knowledge gathered during the last decades [186-195], the following sections illustrate the possibility to explain the neuronal basis of gaze visual orientation, without recourse to notions that belong more to the cybernetic discourse than to the neurophysiology.

## 4.1. Extraocular muscles

Each eye movement involves synergies between anatomo-physiological elements that are invisible to investigations limited to behavioral analyses. However, the latter are helpful for discovering constraints in the contraction changes of the six muscles attached to the surface of each ocular globe (Figure 4). Combined with the relaxation of the medial rectus (MR) muscle fibers, the contraction of the lateral rectus (LR) muscle fibers of the same eye turns the eyeball toward the temporal side. Conversely, the contraction of MR combined with the relaxation of LR rotates the opposite eye toward the nasal side. In both cases, the eye rotates within the plane passing through the two muscles (toward the right in Fig. 4A). These muscles are antagonist to each other because their insertions are symmetrically distributed on opposite sides of the globe [196].

When the head is upright, the eye movement is horizontal provided that no change happens in the balances made by the contraction level of the other four extraocular muscles, between, on the one hand, the pair composed of superior rectus (SR) and inferior oblique (IO) muscles, and on the other hand, the pair composed of the inferior rectus (IR) and superior oblique (SO) muscles (Fig. 5A). Any change in the balance of forces exerted by these two pairs will add a vertical component to the horizontal eye movement. An up-ward deflection of horizontal saccades if the infraductive actions of IR and SO muscle fibers does not counterbalance changes in the contraction of SR and IO muscle fibers. Reciprocally, horizontal saccades will exhibit a downward deflection if the supraductive action of SR and IO does not counterbalance changes in the contraction of IR and SO muscle fibers.

An upward eye movement results from the contraction of SR and IO muscle fibers of both eyes combined with the relaxation of IR and SO of both eyes (Fig. 4B) whereas a downward eye movement results from the contraction of latter combined with the relaxation of the former (Fig. 4C). Their trajectory is straight vertical if the balance between the adduction and abduction does not change.

Anatomo-physiological studies teach us that for each eye, distinct motor and pre-motor cells innervate these different muscles while behavioral studies report horizontal and vertical movements of both eyes with trajectories that are quasi-rectilinear (not curved) and torsionless. Visible to the external observer only, such trajectories imply the existence of synergies in the recruitment of extraocular muscles, synergies that the sole anatomical survey cannot deduce easily.

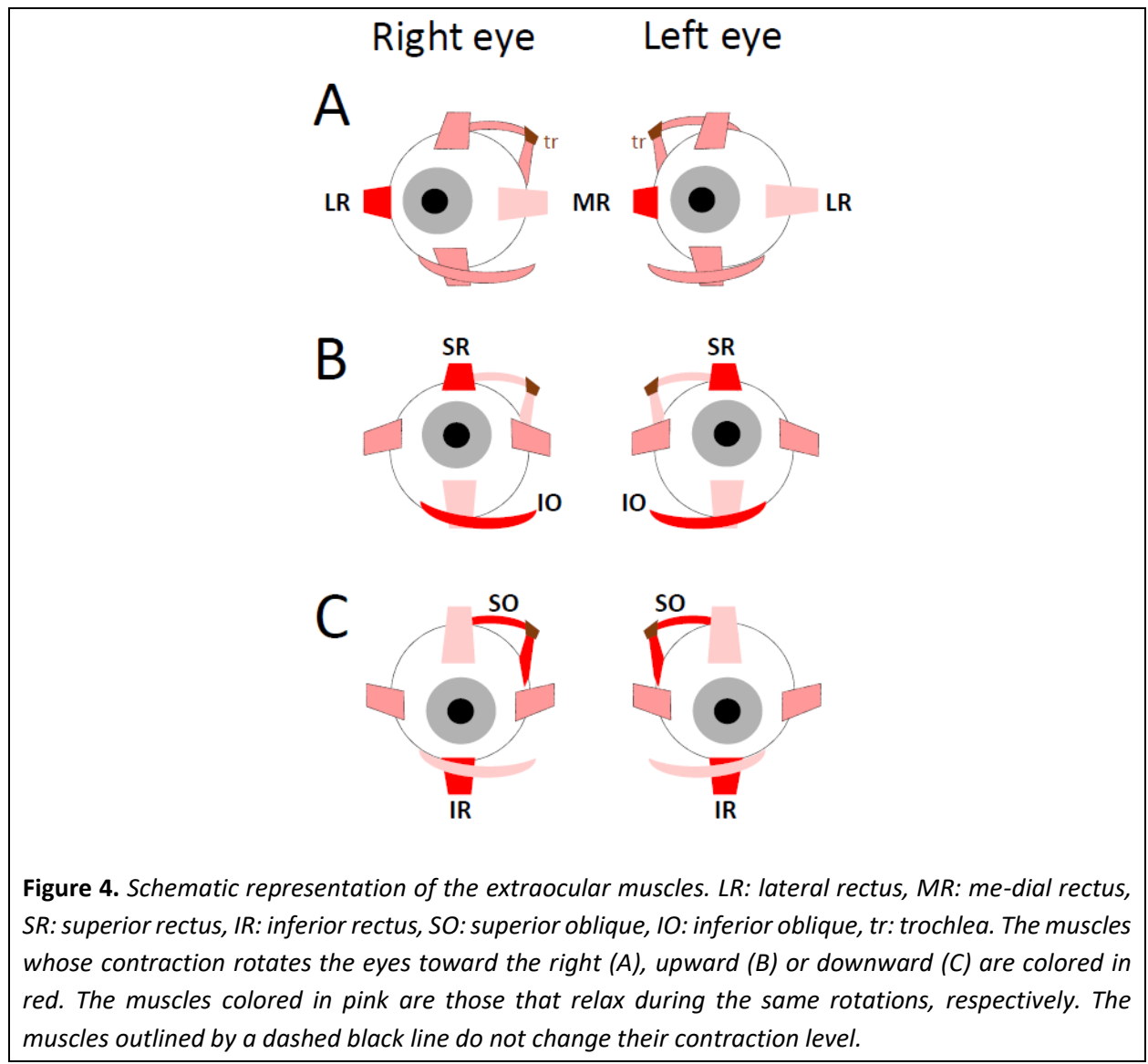

**Figure 4.** *Schematic representation of the extraocular muscles. LR: lateral rectus, MR: me-dial rectus, SR: superior rectus, IR: inferior rectus, SO: superior oblique, IO: inferior oblique, tr: trochlea. The muscles whose contraction rotates the eyes toward the right (A), upward (B) or downward (C) are colored in red. The muscles colored in pink are those that relax during the same rotations, respectively. The muscles outlined by a dashed black line do not change their contraction level.*

## 4.2. Binocular synergy

Orienting movements of gaze toward a visual target involve both eyeballs. The rotation of the left eye results from changes in the contraction of the fibers of its extra-ocular muscles, which are innervated by motor neurons that are distinct from the cells innervating the muscles of the right eye. A synergy is thus required between the groups of premotor neurons for moving both eyes quasi-simultaneously.

### 4.2.1. Horizontal eye movements

Figure 5 illustrates the network responsible for leftward rotations of both ocular globes, i.e., for the contraction of the left eye's LR muscle and the right eye's MR, combined with the relaxation of antagonist muscles (left eye's MR muscle and right eye's LR). As explained above, the contraction of the other four extraocular muscles (SO and IR against IO and SR) is maintained if the movement is straight horizontal (Fig. 5A).

The degree of contraction of an extraocular muscle depends on the frequency of action potentials emitted by motor neurons innervating its fibers. For a horizontal eye movement, the motor cells innervating the LR muscle are grouped in the ipsilateral abducens nucleus whereas those innervating the MR muscle are in the ipsilateral oculomotor nucleus. The abducens nuclei are located in the pontine reticular formation and the oculomotor nuclei in the midbrain (Fig. 5B).

The contraction of agonist muscles is caused by an increase in the firing rate of motor neurons (MN) in the abducens nucleus for the LR muscle and in the oculomotor nucleus for the MR muscle (Fig. 5C). During saccades, the muscle contraction is brisk not only be-cause of the contractile properties of extraocular muscles [188, 190, 196], but also because the incoming input comes in a burst of action potentials with short interspike intervals [197-202]. In the abducens nucleus, internuclear neurons (AIN) relay to the motor neurons that innervate the MR muscle (synapse b) the excitation they receive at the same time as MN from excitatory burst neurons (EBN, synapses a) located in the paramedian pontine reticular formation (PPRF) [201-205].

In the monkey, PPRF lesion eliminates the possibility of producing saccades directed towards the lesioned side (ipsilesional saccades) whereas the generation of slow eye movements (during the vestibular ocular reflex or pursuit) remains possible [118, 206]. Electrical microstimulation applied in the PPRF evokes an ipsilateral movement of both eyes, which does not resemble a saccade. Its speed is most often constant and depends on the stimulation frequency [207-208]. The electrically-evoked movement is not saccade-like for two reasons at least. Firstly, the stimulation parameters (constant current and frequency of pulses) do not replicate the complete population burst that usually precedes each saccade. Indeed, a limitation of the electrical stimulation technique is that its strength cannot be enhanced without exciting unrelated fibers of passage and eliciting reactions other than oculomotor. Secondly, the antagonist muscles were not relaxed as they usually are during saccades. Their relaxation results from a pause in the firing rate of motor neurons, which pause is caused by spikes emitted by inhibitory burst neurons (IBN) located in the contralateral dorsal *paragigantocellularis* reticular formation (dPGRF) [209-212].

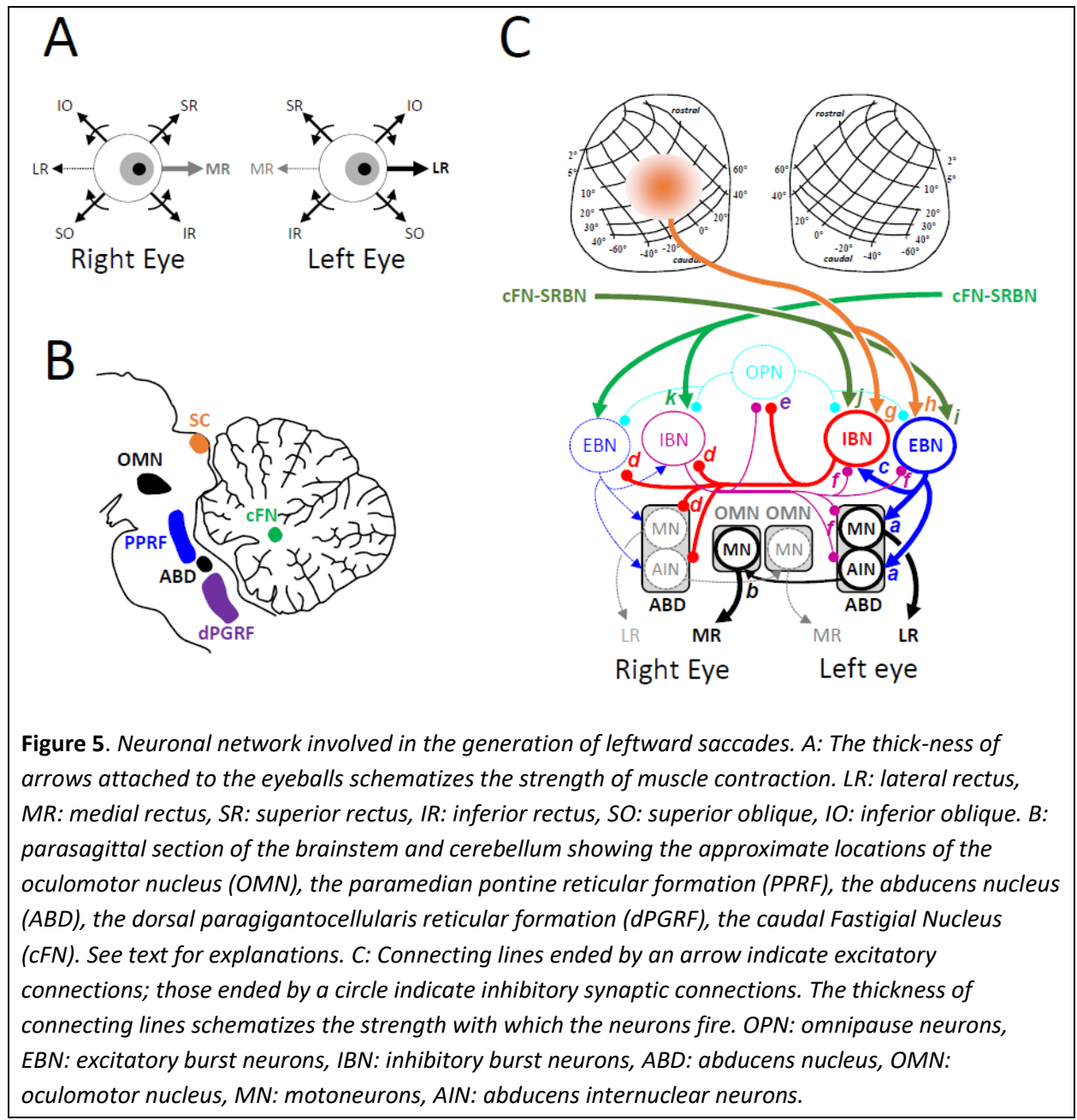

**Figure 5**. *Neuronal network involved in the generation of leftward saccades. A: The thick-ness of arrows attached to the eyeballs schematizes the strength of muscle contraction. LR: lateral rectus, MR: medial rectus, SR: superior rectus, IR: inferior rectus, SO: superior oblique, IO: inferior oblique. B: parasagittal section of the brainstem and cerebellum showing the approximate locations of the oculomotor nucleus (OMN), the paramedian pontine reticular formation (PPRF), the abducens nucleus (ABD), the dorsal paragigantocellularis reticular formation (dPGRF), the caudal Fastigial Nucleus (cFN). See text for explanations. C: Connecting lines ended by an arrow indicate excitatory connections; those ended by a circle indicate inhibitory synaptic connections. The thickness of connecting lines schematizes the strength with which the neurons fire. OPN: omnipause neurons, EBN: excitatory burst neurons, IBN: inhibitory burst neurons, ABD: abducens nucleus, OMN: oculomotor nucleus, MN: motoneurons, AIN: abducens internuclear neurons.*

The consequences of IBN lesion have not yet been studied. However, because of the inhibition they exert on the motoneurons innervating the antagonist muscles and neurons (synapses d), the suppression of their burst is expected to hinder saccades and reduce their speed and amplitude. By disinhibiting IBN located in the opposite side (synapses f), the unilateral suppression of IBN activity is also expected to increase the agonist drive during contralesional saccades and render them hypermetric. Indeed, recordings in the monkey reveal that approximately half of IBN in the side opposite to the recruited EBN emit spikes during contralateral saccades. These action potentials combine with those emitted by EBN (synapses a and f) to compose the agonist oculomotor command [91, 203, 213-215].

Excitatory and inhibitory burst neurons receive descending input from saccade related neurons in the contralateral SC (synapses g and h) and cFN (synapses i and j). The collicular input is monosynaptic in the cat [216] but disynaptic in the monkey [217]. This difference between the feline and primate species also concerns the fastigio-collicular projections. In the monkey, they are concentrated in the rostral part of both SC [79] whereas they target more caudal regions in the cat [218].

Finally, although premotor burst neurons receive lateralized input from the contralateral SC and cFN, the premotor control of saccades should be considered as bilateral [91, 203, 213-215] because of the omnidirectional bursts of saccade-related fastigial neurons. Indeed, for each horizontal saccade, a burst is emitted by neurons not only in the contralateral, but also in the ipsilateral cFN [219-222]. The contralateral burst was proposed to accelerate the saccade by exciting the agonist drive from EBN (synapses i and j) whereas the ipsilateral burst was supposed to decelerate or brake the saccade by recruiting IBN in the opposite side (synapse k).

Several facts led to revise this sequential activation and "biphasic" view of the fastigial control of saccade acceleration and deceleration [223] and to propose the conjecture of bilateral fastigial control of saccades [72, 213-215]. Firstly, the biphasic hypothesis is not applicable to the generation of vertical saccades [72, 215]. Secondly, all IBN do not fire during the late part of contralateral saccades; some of them burst at saccade onset, as EBN do [211-212]. The same remark holds for the burst of saccade related fastigial neurons during ipsilateral saccades [222]. Moreover, the conjecture that the contralateral burst contributes to the acceleration of saccades is refuted by its later occurrence with larger amplitude of contralateral gaze shifts [224]. Yet, the peak velocity of contralateral gaze shifts is significantly reduced during unilateral cFN inactivation [225]. Finally, still after unilateral cFN inactivation, ipsilateral saccades exhibit increased peak velocity and enhanced dis-placement during the accelerating epoch, regardless of whether they are associated with a head movement [225]. Altogether, these observations are not compatible with the biphasic hypothesis [223], but remain compatible with the bilateral conjecture [72, 213-215].

### 4.2.2. Vertical eye movements

Contrary to the unilateral origin of the excitatory drive of binocular horizontal saccades, the excitatory phasic drive is bilateral for saccades directed upward or downward. An upward saccade involves the contraction of SR and IO muscle fibers of both eyes combined with the relaxation of IR and SO muscle fibers, while the contraction of adductive and abductive muscles (LR and MR) does not change (Fig. 6A).

The generation of downward saccade involves the contraction of IR and SO muscle fibers of both eyes combined with the binocular relaxation of SR and IO muscle fibers (Fig. 7A). The contraction of agonist muscles is driven by a burst of spikes emitted by motor neurons located in the ipsilateral oculomotor nucleus (for the IR muscles) and the contralateral trochlear nucleus (for the SO muscles) while the relaxation of antagonist muscles results from a pause in the firing rate of motor neurons located in the ipsilateral oculomotor nucleus (for the IO muscles) and the contralateral one (for the SR muscles).

The contraction of agonist muscles is driven by a burst of action potentials emitted by motor neurons located in the ipsilateral and contralateral oculomotor nuclei (for the IO and SR muscle fibers, respectively) [226-228]. The relaxation of antagonist muscles is permitted by a pause in the firing rate of motor neurons located in the ipsilateral oculomotor nucleus (for the IR muscles) and the contralateral trochlear nucleus (for the SO muscles) (Fig. 6B). The agonist motor neurons receive a bilateral burst command from uEBN located in both (left and right) rostral interstitial nuclei of the medial longitudinal fasciculus (RIMLF; synapses a and b in Fig. 6C) [229-231]. Likewise, the inhibition of antagonist motor neurons is bilateral and driven by uEBN (synapses c). It originates from uIBN located in the contralateral interstitial nuclei of Cajal (INC; synapses d and e).

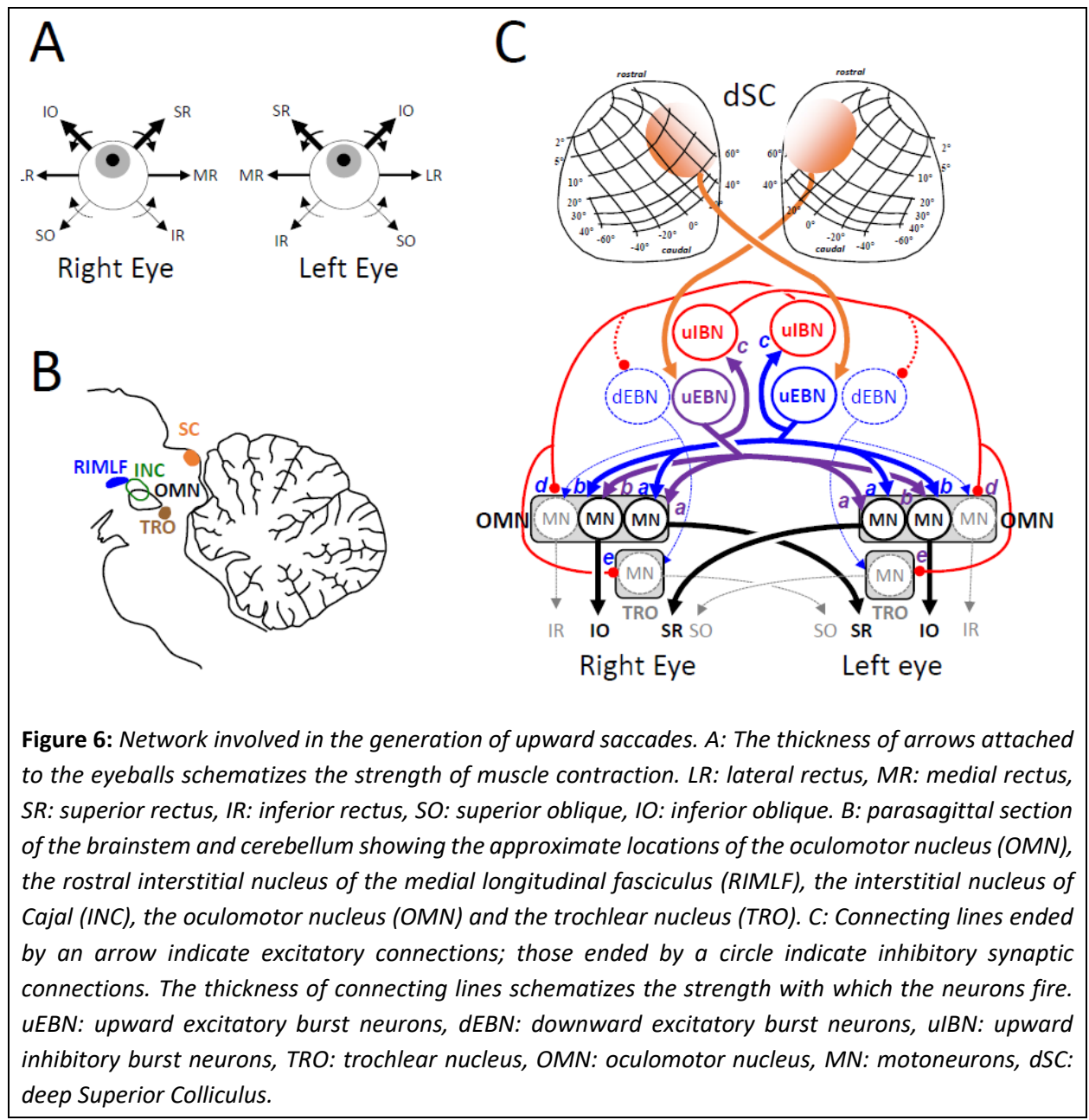

**Figure 6:** *Network involved in the generation of upward saccades. A: The thickness of arrows attached to the eyeballs schematizes the strength of muscle contraction. LR: lateral rectus, MR: medial rectus, SR: superior rectus, IR: inferior rectus, SO: superior oblique, IO: inferior oblique. B: parasagittal section of the brainstem and cerebellum showing the approximate locations of the oculomotor nucleus (OMN), the rostral interstitial nucleus of the medial longitudinal fasciculus (RIMLF), the interstitial nucleus of Cajal (INC), the oculomotor nucleus (OMN) and the trochlear nucleus (TRO). C: Connecting lines ended by an arrow indicate excitatory connections; those ended by a circle indicate inhibitory synaptic connections. The thickness of connecting lines schematizes the strength with which the neurons fire. uEBN: upward excitatory burst neurons, dEBN: downward excitatory burst neurons, uIBN: upward inhibitory burst neurons, TRO: trochlear nucleus, OMN: oculomotor nucleus, MN: motoneurons, dSC: deep Superior Colliculus.*

Contrary to the bilateral origin of the excitatory premotor drive of upward saccades, it seems to be unilateral for downward saccades [232], originating from dEBN located in the ipsilateral riMLF (synapses a and b). Regarding the origin of the inhibition of antagonist motor neurons, it has not yet been identified [233]. The asymmetry of the excitatory input is consistent with the effects of unilateral lesion of riMLF since downward saccades are more impaired than upward ones [234].

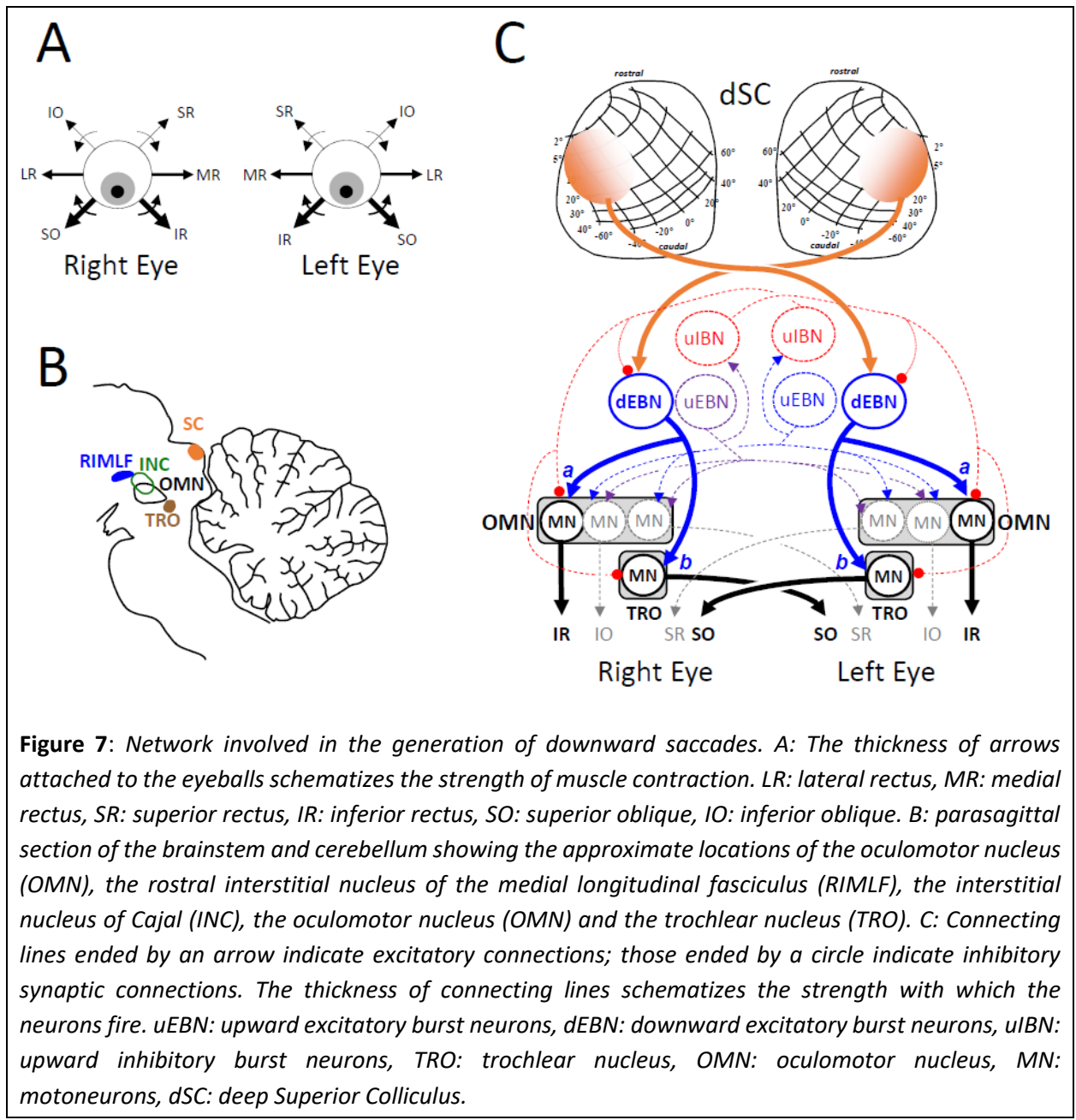

**Figure 7**: *Network involved in the generation of downward saccades. A: The thickness of arrows attached to the eyeballs schematizes the strength of muscle contraction. LR: lateral rectus, MR: medial rectus, SR: superior rectus, IR: inferior rectus, SO: superior oblique, IO: inferior oblique. B: parasagittal section of the brainstem and cerebellum showing the approximate locations of the oculomotor nucleus (OMN), the rostral interstitial nucleus of the medial longitudinal fasciculus (RIMLF), the interstitial nucleus of Cajal (INC), the oculomotor nucleus (OMN) and the trochlear nucleus (TRO). C: Connecting lines ended by an arrow indicate excitatory connections; those ended by a circle indicate inhibitory synaptic connections. The thickness of connecting lines schematizes the strength with which the neurons fire. uEBN: upward excitatory burst neurons, dEBN: downward excitatory burst neurons, uIBN: upward inhibitory burst neurons, TRO: trochlear nucleus, OMN: oculomotor nucleus, MN: motoneurons, dSC: deep Superior Colliculus.*

It is worth noting that i) the contraction of SR and IO muscle fibers does not only supraduct the eye during upward saccades and that ii) the contraction of IR and SO muscle fibers does not only infraduct the eye during downward saccades. The contraction of SR and SO muscle fibers also causes incycloduction of the globe while the contraction of IO and IR muscle fibers causes excycloduction. Thus, during upward saccades, the incycloduction caused by the motor commands sent to the SR combines with the excycloduction caused by the commands to the IO, whereas during downward saccades, the excycloduction caused by spikes exciting the IR combines with the incycloduction caused by the spikes exciting the SO. Behavioral measurements made in healthy subjects reveal negligible torsional component during vertical saccades [235], suggesting that the incycloduction and the excycloduction cancel each other out.

During upward saccades, such a balance must result from bilateral activity in the midbrain because for each eye, the motor neurons innervating the SR muscle are located in the contralateral

OMN whereas the motor neurons innervating the IO muscle are located in the ipsilateral OMN (Fig. 6C). Likewise, during downward saccades, the balance must also result from bilateral activity in the midbrain. The motor neurons innervating the IR muscle are located in the ipsilateral OMN whereas the motor neurons innervating the SO muscle are located in the contralateral trochlear nucleus (Fig. 7C).

Thus, the generation of vertical saccades that are straight and torsionless rests upon adjusted bilateral motor activation, which must take into account both the structural (neurons and their connectivity) and functional (firing properties, secondary and tertiary consequences of muscle contractions) asymmetries existing between oculomotor territories distributed on either side of the brain stem. In the monkey, stimulation of the right riMLF produces a conjugate clockwise rotation of both eyes, whereas left riMLF stimulation produces counter clockwise rotation [236].

Since the generation of vertical saccades is bilateral, a torsional component should appear in vertical saccades after unilateral lesion of riMLF with a direction that depends upon the side of the lesion. When dEBN in the right riMLF are lesioned, activity in the opposite (unimpaired) riMLF should provoke a binocular counter clockwise torsion, i.e., an incyclotorsion of the right eye (caused by SO contraction) associated with an excycloduction of the left eye (caused by IR contraction). Clinical observations made in a patient suffering from a unilateral midbrain lesion affecting the right riMLF confirm this oculomotor disorder. A counter clockwise torsional eye rotation accompanies each voluntary vertical saccades, downward and upward [237]. The fact that the patient's upward saccades exhibited a counter clockwise torsion means that the saccade of the right eye (caused by SR contraction) was associated with an incycloduction while the saccade of the left eye (caused by IO contraction) was associated with an excycloduction.

These clinical observations are important because they suggests that the ipsilateral projection of uEBN toward the oculomotor nucleus is stronger than the contralateral projection. Otherwise, during upward saccades, the counter clockwise torsion would have been counterbalanced by the clockwise torsion (resulting from the IO contraction in the right eye and from the SR contraction in the left eye) promoted by the crossed projections from uEBN to the right OMN. More fundamentally, such inferences illustrate the mutual benefit of combining neurophysiological and neuroanatomical studies in non-human primates with clinical studies in patients for advancing our understanding of oculomotor disorders as well as normal physiology.

Finally, the bilateral control of vertical saccades likely involves the two superior colliculi, with both medial halves driving upward saccades (Fig. 6C) and both lateral halves driving downward saccades (Fig. 7C). In line with the bilateral conjecture mentioned earlier, saccade-related neurons in the midline cerebellum may be involved in compensating for asymmetries and regulating the bilateral balance of activity between the left and right parts of the brainstem. Indeed, most neurons in the caudal fastigial nuclei [219-220] and in the vermis [238-239] emit a burst of spikes during vertical saccades. The paucity of fastigial projections to the midbrain [240] and the numerous studies showing that the asymmetrical functional perturbation of the midline cerebellum primarily impairs the horizontal component of saccades [72] suggest a weak involvement in the generation of vertical and torsional components of saccades. Vertical saccades exhibit a deflection of their trajectory toward the side of caudal fastigial inactivation [72, 241-243] or toward the opposite side when the vermis is asymmetrically lesioned [244-245].

However, further neuroanatomical studies should investigate the cerebellar connectivity with the midbrain because another clinical study reported a transient counter clockwise torsion during horizontal saccades in a patient suffering from a lesion involving the left deep cerebellar nuclei and the left lateral medulla oblongata [246]. Reciprocally, further clinical studies should test whether the bilateral aspect of lesions accounts for the rare changes in Listing's law in patients suffering from cerebellar ataxia [247].

### 4.2.3. Synergy of horizontal and vertical saccades

The burst neurons involved in the generation of horizontal and vertical saccades are located in the pontomedullary reticular formation and in the midbrain, respectively. The pontomedullary reticular formation being more caudal than the midbrain, the distance "traveled" by a collicular action potential to post-synaptic premotor burst neurons differs between horizontal or vertical saccades. Yet, when we examine the velocity profiles of the horizontal and vertical components of oblique saccades, we see that the two components start at the same time. This synchrony may result from releasing the premotor burst neurons from the inhibition exerted by omnipause neurons [248]. Simultaneously released from their tonic inhibition (synapse e in Fig. 5), the premotor burst neurons can emit their burst of action potentials synchronously. According to this conjecture, lesion of RIP should desynchronize the onset of horizontal and vertical components during oblique saccades. No experiment has yet confirmed this prediction.

Another coupling exists between the horizontal and vertical components of oblique saccades. Its neurophysiological substrate has not been elucidated either. When the speed of one component diminishes, the other component is also slowed down. The duration of both component is lengthened during oblique saccades. When one compares cardinal and oblique saccades having equal horizontal displacement amplitudes, the horizontal component is slower during oblique saccades. Thus, the two components are not executed independently of each other. Their interaction, called component stretching, does not seem to originate only from the density of post-synaptic input from different collicular sectors to the populations of burst neurons in PPRF and midbrain [249]. A component stretching even occurs in oblique saccades evoked by an electrical microstimulation of the same collicular site, when the amplitude of one component is artificially varied by triggering the electrically-evoked saccade at different times after a visual saccade [250].

The most spectacular demonstration of component stretching was brought after transient inactivation of neurons in the PPRF [251-252]. Therein, the local injection of lidocaine reduced both horizontal and vertical components of oblique saccades. In some experiments, the inactivation was subtle enough to slow down the saccades without altering their accuracy. Consistent with the hypothesis of a negative feedback control of saccade amplitude, their results also indicate that the elements that have been inactivated do not participate in the feedback that updates the residual motor error. They are also compatible with the hypothesis of poly-equilibrium insofar as the saccade duration is prolonged if the activity of neurons participating in its restoration is suppressed.

### 4.3. The synergy of eye and head movements

Orienting gaze toward the location of a stimulus often involves combined movements of the eyes and head. A change in the orientation of the trunk may also accompany the head movement [8]. When the body is at rest, the head carrying the eyeballs, the gaze movement amplitude corresponds to the sum of the amplitudes of the eyes and head movements. Most of the time, gaze captures the target (i.e., its image is foveated) before the head stops moving because the saccadic eye movement is much faster than the head movement. Then, while the head continues to move, the eyes rotate in the opposite direction. The interval during which the eyes rotate in the direction opposite to the head movement is a reflex response called vestibulo-ocular reflex (VOR). A slow counter-rotation of the eyes in their orbit can also occur before the saccade is launched towards the target, the head rotation preceding that of the eyes as during orienting toward an expected target location [253]. Thus, depending upon the amount of head rotation, the eye saccade will be of greater or lesser amplitude. The amplitude also depends on the deviation of the eyeballs in their orbit. If the eyes are deviated in the direction opposite to the direction of the impending saccade, the saccade amplitude is greater [254-256]; the eyes having a wider range of mobility before reaching the maximum deviation imposed by the oculomotor range. Consequently, the contribution of the head to the gaze displacement amplitude is smaller than during gaze movements initiated when the deviation of the eyes is closer to the limits of the oculomotor range. The head will start moving sooner. Thus, the contribution of the head to the gaze displacement amplitude depends also upon the delay between the onset of its movement and the saccade onset. Regardless of whether the head moves or not, small gaze movements are equally accurate [257-258]. For targets requiring large gaze shifts, the accuracy is improved when a head movement accompanies the eye saccade [257].

In the macaque, changing the orientation of the head involves at least twenty muscles [259]. Three groups can be distinguished: i) the groups of muscles whose contraction changes the orientation of the head with respect to the trunk, ii) the "orthogonal" groups of muscles whose contraction stabilizes the atlas while it rotates under the influence of the first group, and iii) the group of muscles that do no change their contraction level [260-261]. For instance, the co-contraction of the two *rectus capitis posterior minor* (RCPm) muscles stabilizes the atlas while the contraction of the rectus *capitis posterior major* (RCPM) and *obliquus capitis inferior* (OCI) muscles produces an ipsilateral horizontal rotation. The contralateral RCPM and OCI muscles show no change in electromyographic (EMG) activity. This lack of change in the activity of antagonist muscles indicates that stopping the head rotation does not involve their contraction. For movements of amplitude greater than 20 degrees, the contraction of the *splenius capitis* muscle completes the contraction of RCPM and OCI muscles. The co-contraction of the RCPm, RCPM, OCI and *obliquus capitis superior* (OCS) muscles stabilizes the head at the end of its change in orientation.

In summary, when it involves a head movement, the gaze orienting response mobilizes several muscles, the synergy of which results from an activity involving a complex set of motor and premotor neurons, in the reticular formation, the vestibular nuclei and the spinal cord. In the continuation of studies performed in the cat, neurophysiological investigations of eye-head movements in monkeys boomed in the 1990s and the beginning of the 21st century. They were dominated by the assumption that gaze direction, i.e., the direction of the line of sight in space was an intrinsic cerebral command controlling eye and head premotor systems located downstream (for an alternative view see [262]). Three major obstacles hampered the development of experimental investigations on simian eye-head movements. The first one was the technical difficulty of carrying out experiments with monkeys whose head is free to move. Added to this was the challenge posed by the need to merge the new neurophysiological data collected in non-human primates with the considerable body of knowledge that had been obtained during numerous experiments with the feline species. Differences between the feline and primate species started to be noticed, questioning the relevance of data collected in non-primate animals to understand the physiology of the human brain. Thirdly, the attractiveness of this developing field of research in macaque monkeys was destroyed by political factors that promoted fast publishing research works as well as the development of brain imaging and in-computer modeling studies. Faced with a migration of young investigators towards scientific fields in which the work was easier or more fun to perform, the neurophysiological study of eye-head coupling in monkeys declined. Societal changes in western countries complicated and rendered socially thankless the use of non-human primates for the advancement of scientific knowledge, discouraging the few researchers who remained engaged.

Fortunately, past results obtained in the cat left a treasure that allows a synthesis, which, although incomplete, illustrates the tremendous complexity underlying the measured changes in the orientations of the eyes and head. The following text completes a re-cent report [91] with more detailed explanations and relevant references. It will serve as a guide for further investigations with marmosets or for comparison with other species.

### 4.3.1. Reticulo-vestibulo-reticular synergies

In the cat, the recruitment of neck muscles during horizontal head movements exhibits some peculiarities. Although some muscles are systematically recruited, either phasically (splenius muscles, OCI, *levator scapulae and complexus*) or tonically (*biventer cervicis*), the recruitment of other muscles (*semispinalis cervicis*, *longissimus, levator scapulae*, *scalenus anterior* and OCS) depends on the orientation of the head relative to the trunk [263]. At the neurophysiological level, the motor network responsible for horizontal movements can be divided into three groups: an excitatory medial reticulospinal system and two medial vestibulospinal subsystems, a contralateral excitatory subsystem and an ipsilateral inhibitory subsystem. Contrary to the macaque, almost every vestibulospinal neuron projects to the ocular motor nuclei in the cat. In the monkey, the relative absence of cells carrying the same signals to motor neurons innervating the extraocular and neck muscles accounts for the dominance of the vestibulo-ocular reflex over the vestibulo-collic [264].

#### 4.3.1.1. The reticulospinal channel

Motor neurons that innervate the agonist muscles receive signals from neurons lo-cated ipsilaterally in the rostrodorsal portion of the nucleus *reticularis gigantocellularis* (NRGC), the nucleus *reticularis pontis caudalis* (NRPC), and the dorsal portion of the nucleus *reticularis gigantocellularis* [265]. The combined lesion of the NRPC and NRGC abolishes almost completely the generation of ipsilateral head movements [266]. Ipsilesional saccadic movements of the eyes are also absent. However, contralateral saccades and vertical movements of the eyes and head do not seem to be affected. The slow compensatory movements of the vestibulo-ocular and vestibulo-collic reflexes are also spared.

The firing properties and connectivity of reticulospinal neurons are compatible with their involvement in the generation of combined ipsilateral eye and head movements [267]. Indeed, eye-neck reticulospinal neurons (EN-RSN) emit bursts of action potentials during ipsilateral eye saccades associated with ipsilateral EMG activity of neck muscles. When the eyes are deviated toward the contralateral side and the ipsilateral neck muscles are relaxed, the neurons are silent. Action potentials are emitted only when the saccade leads the eyes beyond the orbital sagittal plane. None is emitted when the saccade is made in the contralateral orbital hemifield. This sensitivity to the orbital eye deviation, which appears neither in tecto-reticulo-spinal neurons [268-269] nor in cortico-reticulo-spinal neurons [270], may result from afferents from neurons in the nucleus *prepositus hypoglossi* (NPH). Therein, neurons evince a sustained firing rate that increases with the ipsilateral deviation of the eyes in the orbit. When the eyes are deviated toward the contralateral side, they are silent [271]. Thus, NPH neurons may account for silencing the EN-RSNs because their projection are primarily contralateral and inhibitory [91].

A projection of omnipause neurons onto the EN-RSN has been interpreted as evidence for an involvement of OPN not only in the generation of eye saccades but also in eye-head gaze saccades [272]. This conjecture is compatible with a stronger correlation between the duration of their pause and the duration of gaze saccades rather than with the duration of eye saccades [273]. It is also compatible with the slowing of head movement when a short electrical microstimulation is applied in the RIP nucleus [273]. However, the alternative interpretation that OPN remain primarily involved in the generation of ocular saccades nevertheless cannot be rejected. The fact that the eye saccade duration is shorter than the gaze saccade duration does not necessarily imply that the burst duration of pre-motor neurons is also shorter. The shorter duration of eye saccades may result from the action of the vestibulo-ocular reflex downstream, at the level of motor neurons. Thus, the correlation between the pause duration of OPN and the duration of eye saccades would be less strong than the correlation with the duration of gaze saccades. Regarding the slowing of head movement by the electrical microstimulation of OPN region, it may result from a transient re-activation of the vestibulo-collic and vestibulo-ocular reflexes because of the inhibition that OPN exert upon the excitatory and inhibitory burst neurons [69]. Finally, it is worth signaling that in the monkey, the slowing of head movement is rarely observed when the OPN region is stimulated [22] and that the end of the OPN pause is better correlated with eye saccade end than with gaze shift end [274].

### 4.3.1.2. The vestibulospinal channels

The medial vestibulospinal tract (MVST) is the major pathway by which afferents from the horizontal semicircular canals influence the axial musculature of the cat's body. A tri-synaptic pathway connects the labyrinth to the motor neurons innervating the neck muscles [275-276]. The neurons at the origin of the vestibulo-nucal pathway are secondary vestibular neurons located in the medial vestibular nucleus (MVN). Activated by ipsilateral vestibular afferents, the firing rate of these neurons increases during an ipsilateral rotation of the head and decreases during a contralateral rotation, a response called of type I [277-278]. The post-synaptic influence of the MVST onto the motor neurons of the contralateral cervical cord is excitatory. An inhibitory monosynaptic projection from the MVN to ipsilateral neck motoneurons has also been shown [276-277, 279]. The fact that a stimulation of the horizontal canal evokes a head movement directed towards the contralateral side [280] is compatible with this organization. Many neurons of the MVST, mainly located in the rostral region of the MVN, emit a collateral towards the contralateral abducens nucleus [281-283]. This double connection with the extraocular and head motor neurons demonstrates an involvement in combined movements of the eyes and the head.

Electrophysiological studies report relatively high sustained activity of Eye-Neck Vestibular neurons when the eyes are centered in the orbit [284]. This tonic firing rate results either from an intrinsic property of neurons or from the regular excitation from primary vestibular afferents [285]. In some vestibular neurons, a sensitivity of their firing rate to the deviation of the head relative to the trunk suggests an influence also from proprioceptive afferents from neck muscles [286].

During a passive rotation of the head, the vestibular signals inhibit the motor neurons innervating the ipsilateral extraocular and neck muscles while facilitating the activity of those innervating the contralateral muscles. Thus, the rotation of the head promote the emission of vestibular signals that lead to the counter-rotation of the eyes (vestibulo-ocular reflex) and of the head (vestibulo-collic reflex) in the direction opposite to the head rotation [see Fig. 6 in [91]]. These reflex movements also oppose against active rotations of the head. We shall now see how saccade-related commands remove this opposition.

### 4.3.2. Inhibition of vestibulo-ocular and vestibulo-collic reflexes

In addition to excitatory input from EBN and EN-RSN (synapses b), the motor neurons innervating the agonist eye and neck muscles (E-MN and N-MN, respectively) also receive inhibitory afferents from the IVN-1 located in the ipsilateral MVN (synapses c) and excitatory afferents from the EVN-1 in the contralateral MVN (synapses d). The inhibition of agonist motor neurons (by IVN-1, synapse c)) is prevented by spikes emitted by IVN-2 (synapse e), which we just saw are recruited by EN-RSN and EBN (synapses a). It is also prevented by spikes emitted by EVN-1 located in the contralateral MVN (synapses f). The contralateral EVN-1 are indeed released from the inhibition of IVN-2 (synapse g) by the burst emitted by IBN (synapse h), which are recruited by agonist premotor neurons (synapses i). Thus, a disinhibition facilitates the recruitment of agonist muscles while the relaxation of antagonist muscles is promoted by the inhibition that IBN exerts on IVN-2 in the contralateral MVN (synapse h) and on antagonist ocular premotor and motor neurons (synapses j).

During the combined movement of the eyes and head, the eyes do not counter-rotate (vestibulo-ocular reflex) because the motor and internuclear neurons (E-MN) in the contralateral abducens nucleus are inhibited by the IBN (synapse j). By exciting the IVN-2 (synapses a), the bursts emitted by RSN and EBN prevent the ipsilateral IVN-1 from thwarting the recruitment of agonist motor neurons (synapses c). The counter-rotation of the eyes restarts as the firing rate of ipsilateral EBN and IBN diminishes. The declining firing rate of EBN removes the inhibition that ipsilateral IVN-2 exert on IVN-1 and EVN-1 (synapses e and k) while declining firing rate of IBN disinhibits IVN-2 in the contralateral side (synapse h), which start inhibiting the EVN-1 and IVN-2 (synapses g and l). The ipsilateral MNs are inhibited by the IVN-1 (synapse c) while the contralateral MN can emit again action potentials in proportion to the excitation they receive from the contralateral EVN-1 (synapses m) during the end of the head movement. Since this excitation recruits the antagonist neck muscles, it contributes to stopping the head movement [287-288]. However, it is worth reminding that in the monkey, the co-contraction has been questioned by EMG studies [259-260].

In summary, the firing rate reduction of IVN-2 contributes to end the orienting gaze shift by two parallel processes. The first process is the consequence of the gradual decline of excitation from EBN and EN-RSN, while the second process results from an inhibition by IBN in the contralateral side. The decline of EBN and EN-RSN firing rate might not result from a reduced drive from collicular neurons because the latter often continue to fire after saccade end [289-295]. Instead of a reduced drive, a desynchronization of presynaptic input to premotor neurons may happen. The decline of EBN and EN-RSN activity may result from the inhibitory input from IBN in the contralateral side. Thus, the action potentials emitted by IBN during contralateral saccades would participate in the inhibition of contralateral IVN-2 and promote the counter-rotation of the eyes.

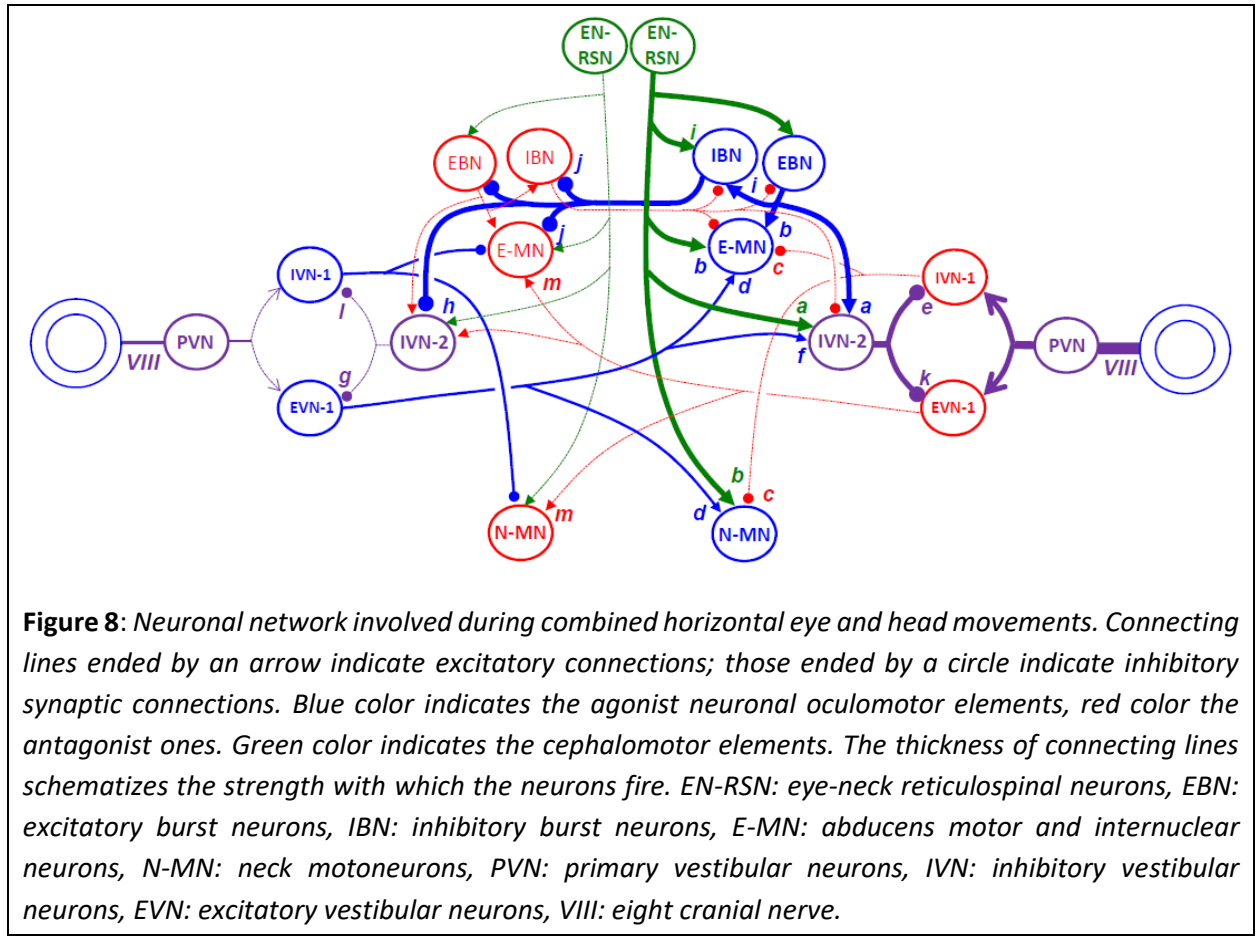

**Figure 8**: *Neuronal network involved during combined horizontal eye and head movements. Connecting lines ended by an arrow indicate excitatory connections; those ended by a circle indicate inhibitory synaptic connections. Blue color indicates the agonist neuronal oculomotor elements, red color the antagonist ones. Green color indicates the cephalomotor elements. The thickness of connecting lines schematizes the strength with which the neurons fire. EN-RSN: eye-neck reticulospinal neurons, EBN: excitatory burst neurons, IBN: inhibitory burst neurons, E-MN: abducens motor and internuclear neurons, N-MN: neck motoneurons, PVN: primary vestibular neurons, IVN: inhibitory vestibular neurons, EVN: excitatory vestibular neurons, VIII: eight cranial nerve.*

# 5. Conclusions

Orienting gaze movements enable an entire animal to localize and capture an object in its physical environment. They are a convenient means for investigating how networks of neurons govern oculomotor behavior [296] and underlie behavioral performances, notably when neurological damage alters their normal functioning [179-180]. During the last decades, following the steps of predecessors who conducted their research in the feline species, neurophysiologists and neuroanatomists gathered in the monkey considerable knowledge that contributed to identify the core networks involved in the generation of eye and head movements [91] and to understand oculomotor disorders in human patients [184-185].

Technological progress brought the possibility to measure precisely the time course of eye and head movements and to study correlations between the firing rate of neurons and various kinematic parameters (amplitude, velocity and acceleration) or the angular distance between gaze and target directions. Analytical habits resulted, which became conventions that gradually led some neuroscientists to implicitly assume a one-to-one correspondence relation between measured physical values of gaze or head orientation and neuronal activity. From there, the search for neurophysiological evidence supporting their theoretical diagram was promoted.

Conflating kinematic or dynamic numerical values with neurophysiological recordings carries the risk of believing that central neuron activity directly encodes gaze or head orientation rather than mediating changes in extraocular and neck muscle contraction. Moreover, neuroanatomical and neurophysiological studies reveal functional properties and connectivity that are not visible in the cybernetic diagrams. The comparison between a diagram and the brain is analog to that between an automaton and a living animal. As early as in 1971, David A. Robinson, one of the founders of the modern neurophysiology of eye movements warned that "block diagrams of oculomotor organization serve as a compact description of system behavior but seldom have much bearing on the way in which the real system, composed of nerve and muscle, actually operates. The models thus do not contribute much to the neurophysiology (or neurology) of eye movements and incur the danger of suggesting that there actually are segregated portions of the nervous system which perform the differentiation, integration and other operations indicated in the boxes of the diagrams" [95]. These reservations do not concern only the operations indicated in the diagrams. They also concern the signals transmitted between the functional blocks.

Complementing a previous synthesis that was inevitably incomplete [91], this review article is enriched by epistemological comments and numerous additional precisions and references. It also provides the solid ground of evidence for a new departure in the neuro-physiological study of orienting movements. Rather than reducing position or error signals - which are numerical values belonging to the physical domain of kinematics -, orienting movements of the eyes and head are the behavioral expressions of intrinsic neuronal processes restoring a poly-equilibrium, i.e., multiple balances of neuronal activities opposing mutually antagonistic visuomotor channels.

## Acknowledgments:

This research was funded by Centre National de la Recherche Scientifique (CNRS). The author thanks Dr Blas Torres and Dr David L. Sparks for accepting the inclusion of one of their figure in this synthesis article.